# Angle-dependent resonant tunneling and thermoelectric energy management in a hybrid 1D-2D-1D semiconductor nanostructure

Xiaoguang Luo[1,*], Jiaming Wang[1], Jiawen Dai[1,2], Junqiang Zhang[1], and Nian Liu[3]

[1] State Key Laboratory of Flexible Electronics & Institute of Flexible Electronics, Northwestern Polytechnical University, Xi'an 710129, China

[2] School of Mechanics, Civil Engineering and Architecture, Northwestern Polytechnical University, Xi'an 710129, China

[3] School of Electrical and Electronic Engineering, Anhui Science and Technology University, Bengbu 233100, China

**ABSTRACT**

Low-dimensional semiconductors have been widely exploited in thermoelectric energy conversion for high efficiencies due to their suppressed lattice thermal conduction, sharply defined electronic density of states, and tunable energy-selective electron transmission. However, the widespread challenge of Fermi-level pinning or doping constraints limit precise control over thermoelectric energy management via chemical potential modulation. Here, we proposed an alternative strategy: leveraging angle-dependent electron incidence to dynamically manipulate electron transmission and heat transport, which was implemented theoretically in a two-dimensional InP/InAs/InP double-barrier heterostructure integrated with laterally one-dimensional electrodes. By combining the transfer matrix method and Landauer formalism, we demonstrated the angle-dependent resonant tunneling dynamics, tunable negative differential resistance effect, and near-Carnot limits in thermoelectric energy conversions. Angular modulation enables precise control over transmission resonances, facilitating dynamic transitions among thermoelectric regimes (power generation, cooling, and hybrid heating) without requiring extreme chemical potential shifts. This work establishes angularly resolved electron transmission as a versatile mechanism for on-chip thermal management and cryogenic applications, offering a pathway to circumvent material limitations in next-generation nanoelectronics and quantum devices.

---

* Contact author: iamxgluo@nwpu.edu.cn

## I. INTRODUCTION

Thermoelectric energy conversion has garnered significant interest in recent decades due to its inherent reliability, flexible scalability, environmental sustainability, and bidirectional energy conversion capability, positioning it as a promising technology for addressing global energy challenges[1, 2]. The performance of thermoelectric materials is quantified by the dimensionless figure of merit $ZT = S^2\sigma T/(\kappa_e + \kappa_l)$[3], where $S$ denotes the Seebeck coefficient (thermopower), $\sigma$ the electrical conductivity, $\kappa_e$ and $\kappa_l$ the electronic and lattice thermal conductivities, respectively. Optimal thermoelectric performance commonly requires a delicate balance of high electrical conduction and suppressed thermal transport.

Nanostructured materials have emerged as promising candidates for higher efficiencies than their bulk counterparts. At reduced dimensions, increased interface density on the length scale comparable to the phonon mean-free path, can be used to reduce parasitic lattice thermal conduction[4, 5]. Concurrently, quantum confinement and band engineering enable sharp features in electronic density of states (DOS), enhancing thermopower $S$ and power factor $S^2\sigma$[6, 7]. Theoretical studies suggest that a delta-function-like DOS maximizes $ZT$[8], enabling near-Carnot efficiencies under reversible operating conditions[8, 9]. Practical implementations, however, rely on the precise energy-selective filtering of transmitted electrons[10-12] via tunable energy levels of nanostructures (or the electronic band structure), through doping, electrostatic gating, and external field. These principles have driven advancements in quantum dot[13, 14], nanowire[15], symmetrically-doped nanowire[16], superlattice[17, 18], and core-superlattice[19, 20], which exhibit tailored transport properties for high-efficiency thermoelectric applications.

Semiconductor superlattices–periodic heterostructures of alternating ultrathin layers–offer precise control over DOS and electronic band structures through engineered periodicity. Many semiconductor superlattices with finite periodicity have already been employed in thermoelectric, such as PbSe/PbS[21], Bi$_2$Te$_3$/Sb$_2$Te$_3$[22], GaAs/AlGaAs[23], and InP/InAs[13] systems. At fixed size parameters, their energy-selective transmission properties are determined by electron effective mass, potential profile, and chemical potential, where the latter two factors can be easily modulated via applied electric field. However, practical limitations such as Fermi-level pinning and doping constraints in some semiconductors restrict chemical potential modulation, necessitating alternative strategies for thermoelectric energy management. Recent insights into angle dependent electron transmission[24, 25], governed by Snell' law analogies in superlattices, provide a new avenue for tuning transport properties. Realizing such angular control requires ultrafine one-dimensional (1D) electrodes (e.g., metallic nanowire[26] or nanotube[27]) to establish precise point contacts with superlattice structures.

In this study, we investigated a 1D-2D-1D thermoelectric architecture featuring a 2D InP/InAs/InP double-barrier heterostructure (DBH) as an angle-dependent energy-selective electron filter. Using transfer matrix method (TMM) and Landauer formalism, we analyzed the angle-dependent band structure, current-voltage characteristics, and thermal transport dynamics. Furthermore, we demonstrated the emergence of angle-dependent negative differential resistance (NDR) effects and effective thermoelectric energy management.

## II. ELECTRON TRANSMISSION IN THE 1D-2D-1D SYSTEM

The proposed system employs a 2D InP/InAs/InP DBH as an electron transport medium and energy filter, as shown in Fig. 1(a). Two laterally integrated ultrafine metallic nanowires serve as 1D electrodes for electron injections, with ultrathin InAs buffer layers mitigating complex interfacial effects between 2D and 1D subsystems. The electron effective masses in InAs and InP are $0.023m_0$ and $0.08m_0$, respectively, where $m_0$ denotes the free electron mass in vacuum[28]. Within the DBH, the InP and InAs layers constitute the barrier and quantum well regions, respectively, with a barrier height $\phi = 0.57$ eV[29]. For simplicity, the potentials of InAs and InP are defined as $V_{\text{InAs}} = 0$ eV and $V_{\text{InP}} = 0.57$ eV, such that the total energy $E$ of electrons within the InAs well corresponds to their kinetic energy. Due to the non-collinear alignment of two 1D electrodes along the *x*-direction, electron transmission through the DBH exhibits an angular dependence on incidence angle $\theta$. This 1D-2D-1D system in this study is simplified via a modified Kronig-Penney model, as shown in Fig. 1(b), where $d_b$ (InP layer thickness) and $d_w$ (InAs layer thickness) are fixed at 4 nm and 5 nm, respectively, in the following calculations, unless otherwise specified.

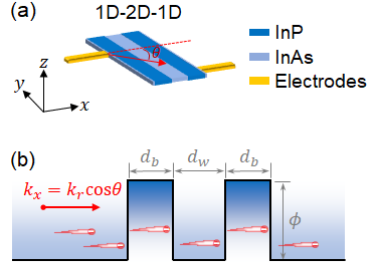

**FIG. 1** (a) The schematic diagram of the 1D-2D-1D structure. (b) Potential profile of the structure along $x$-direction without bias.

At the equilibrium state, electron dynamics within the DBH are governed by the 2D stationary Schrödinger equation:

$$-\frac{\hbar^2}{2m^*(x,y)}\left(\frac{\partial^2}{\partial x^2} + \frac{\partial^2}{\partial y^2}\right)\psi(x,y) + V(x,y)\psi(x,y) = E\psi(x,y) \quad (1)$$

where $\psi(x,y)$, $V(x,y)$ and $m^*(x,y)$ denote the electron wavefunction, potential and position-dependent effective mass, respectively. Owing to translational symmetry along the $y$-direction, the $y$-component of electron wavevector ($k_y$) is conserved across interfaces, as dictated by the electron analogue of Snell's law[24]. Therefore, the potential and effective mass reduce to $V(x)$ and $m^*(x)$, and the wavefunction separates as $\psi(x,y) = \varphi(x)e^{ik_y y}$. The total wavevector ($k_r$) satisfies $k_r^2 = \frac{2m^*(x)[E-V(x)]}{\hbar^2} = k_x^2 + k_y^2$ in each layer.

The general solution to Eq. (1), obtained using the separation of variables approach, is:

$$\psi(x,y) = \left(ae^{ik_x x} + be^{-ik_x x}\right)e^{ik_y y} \quad (2)$$

where $a$ and $b$ are coefficients determined by boundary conditions. The terms $ae^{ik_x x}$ and $be^{-ik_x x}$ in the wavefunction correspond to the right- and left-traveling waves along $x$-direction, respectively[30]. $k_y = k_r \sin\theta$ is determined in the InAs well. The conservation of $k_y$ across the DBH ensures a uniform probability density along $y$-component, reflected in the normalization $\varphi(y)\varphi^*(y) \equiv 1$. Therefore, $a$ and $b$ can be resolved by boundary conditions in $x$-direction, i.e., both the wavefunction $\varphi(x)$ and the probability current density $\frac{1}{m^*(x)}\frac{d\varphi(x)}{dx}$ should be continuous at each InP/InAs interface.

The wavefunction along the propagation direction can be reformulated into a transfer matrix representation to facilitate calculational analysis. This transformation, termed the TMM, is widely employed in calculating the transmission and reflection coefficients for optical[31], acoustic[32], and electronic[17, 33] waves in heterostructures. In the TMM framework, the arbitrary potential profile is discretized into $N$ segments, with finer discretization ($N \gg 1$) enhancing numerical accuracy. In the $j$th segment ($j = 1, 2, \ldots, N$), the general solution of $x$-component wavefunction is: $\varphi_j(x) = A_j + B_j = a_j e^{ik_j x} + b_j e^{-ik_j x}$, which can be rearranged as the vector form:

$$\begin{pmatrix} A_j \\ B_j \end{pmatrix} = \begin{pmatrix} a_j e^{ik_j x} \\ b_j e^{-ik_j x} \end{pmatrix} \quad (3)$$

For brevity, the subscript of "$x$" in the wavevector component $k_x$ is omitted hereafter, unless otherwise specified. Continuity of the wavefunction and its probability current density at interfaces dictates the transfer matrix $T_{j,j+1}$ from segment $j$ to segment $j+1$[17, 33]:

$$T_{j,j+1} = \frac{1}{2} \begin{pmatrix} 1 + \frac{k_{j+1} m_j^*}{k_j m_{j+1}^*} & 1 - \frac{k_{j+1} m_j^*}{k_j m_{j+1}^*} \\ 1 - \frac{k_{j+1} m_j^*}{k_j m_{j+1}^*} & 1 + \frac{k_{j+1} m_j^*}{k_j m_{j+1}^*} \end{pmatrix} \quad (4)$$

and the propagation matrix $P_j$ within a homogeneous segment $j$:

$$P_j = \begin{pmatrix} e^{-ik_j d_j} & 0 \\ 0 & e^{ik_j d_j} \end{pmatrix} \quad (5)$$

Here, the wavevector $k_j$, effective mass $m_j^*$ and thickness $d_j$ are determined by the material in segment $j$. The total transfer matrix $M$ for the entire DBH is constructed as:

$$M = \begin{pmatrix} M_{11} & M_{12} \\ M_{21} & M_{22} \end{pmatrix} = T_{0,1} \cdot P_1 \cdots T_{j-1,j} \cdot P_j T_{j,j+1} \cdots P_N \cdot T_{N,N+1} \quad (6)$$

relating to the incident $(A_0, B_0)$ and transmitted $(A_{N+1}, 0)$ wavefunction via:

$$\begin{pmatrix} A_0 \\ B_0 \end{pmatrix} = M \cdot \begin{pmatrix} A_{N+1} \\ 0 \end{pmatrix} \quad (7)$$

Among them, the subscripts of "0" and "$N+1$" correspond to the uniform media for electron injection and exit.

The electron transmission probability, defined as the ratio of transmitted and incident probability current density, is expressed as:

$$\tau(E, \theta) = \frac{m_0^*}{m_{N+1}^*} \frac{k_{N+1}}{k_0} \left| \frac{1}{M_{11}} \right|^2 \quad (8)$$

where $m_0^*$ and $m_{N+1}^*$ denote the electron effective masses, and $k_0$ and $k_{N+1}$ represent the wavevectors in the incident ($j = 0$) and exit ($j = N+1$) regions of the DBH, respectively. The term $M_{11}$ corresponds to the (1,1) element of the total transfer matrix $M$ defined in Eq. (6), reflecting the cumulative effect of wave interference across the discretized potential profile.

Under zero bias, symmetry dictates $m_0^* = m_{N+1}^* = m_{\text{InAs}}^*$ and $k_0 = k_{N+1}$, simplifying the transmission probability to $\tau(E, \theta) = 1/|M_{11}|^2$. The DBH in our model is equivalent to a two-period counterpart of the InP/InAs superlattice[33]. For such a system, the heterostructure is partitioned into three segments (one material layer per segment) for the transfer matrix analysis. The transmission spectrum of the DBH should align with the electronic band structure of the infinite superlattice. The band structure is derived via the transfer matrix of a single InP/InAs superlattice unit cell, combined with the Bloch's theorem:

$$\varphi(x+d) = e^{i\kappa d}\varphi(x) \tag{9}$$

where $d = d_{InP} + d_{InAs}$ is the lattice constant and $\kappa$ is the Bloch wavevector along $x$-direction. Solving eigenvalue equation yields:

$$\cos(\kappa d) = \cos(k_{InP}d_{InP})\cos(k_{InAs}d_{InAs})$$
$$-\frac{1}{2}\left(\frac{k_{InAs}m_{InP}^*}{k_{InP}m_{InAs}^*} + \frac{k_{InP}m_{InAs}^*}{k_{InAs}m_{InP}^*}\right)\sin(k_{InP}d_{InP})\sin(k_{InAs}d_{InAs}) \tag{10}$$

with the total wavevectors of $|k_r| = \sqrt{2m_{InAs}^*E}/\hbar$ and $|k_r| = \sqrt{2m_{InP}^*(E-\phi)}/\hbar$ in InAs well ($V_{InAs} = 0$) and InP barrier ($V_{InP} = \phi$), respectively. Subsequently, the $x$-components of $k_{InAs}$ and $k_{InP}$ can be calculated from $k_x^2 = k_r^2 - k_y^2$, where $k_y = \sin\theta\sqrt{2m_{InAs}^*E}/\hbar$.

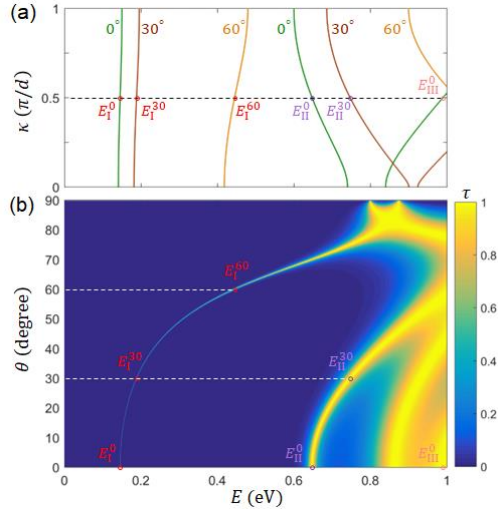

FIG. 2 Electronic band structure of the InP/InAs superlattice and transmission spectrum of the DBH. (a) Calculated band structure of the InP/InAs superlattice for layer thicknesses $d_{InP} = 4$ nm and $d_{InAs} = 5$ nm. Results are shown for incidence angles $\theta = 0°$ (green), 30° (brown) and 60° (orange). (b) The electron transmission spectrum through the DBH for the incidence angle from 0° to 90°. Marked circles indicate peak energies of resonant transmission for the incidence angle at 0°, 30° and 60°, respectively (i.e, first transmission band: $E_I^0$, $E_I^{30}$, and $E_I^{60}$; second transmission band: $E_{II}^0$ and $E_{II}^{30}$; third transmission band: $E_{III}^0$).

Using Eq. (10), the electronic band structures of the InP/InAs superlattice were computed for varying incidence angles. Fig. 2(a) presents the calculated results within the energy range $E < 1.0$ eV for $\theta = 0°$, 30°, and 60°. In an infinite superlattice, perfect transmission ($\tau = 1$) occurs within each miniband at resonant modes where the accumulated phase shift equals integer multiples of $\pi$. For a finite superlattice with $n$-periodicity, the resonant condition is modified to:

$$\kappa d = (m - 1 + \gamma/n)\pi \qquad (11)$$

where $\gamma = 1, 2, 3, \cdots, n-1$ denotes the resonance index within $m$th Brillouin zone[33]. For the two-period DBH, resonant transmission arises at $\kappa d = \pi/2$ in the reduced Brillouin zone. In the energy region of $E < 1.0$ eV, three minibands are identified for $\theta = 0°$ and $30°$, while two minibands emerge for $\theta = 60°$. Resonant states are labeled as $E_I^0$, $E_I^{30}$, and $E_I^{60}$ (first miniband); $E_{II}^0$ and $E_{II}^{30}$ (second miniband); and $E_{III}^0$ (third miniband) in Fig. 2(a). To validate these resonant states, transmission probabilities were calculated using Eq. (8), yielding the angularly resolved spectrum in Fig. 2(b). Transmission peaks with $\tau = 1$ form distinct angular bands that consistent to the electronic band structure in Fig. 2(a). Specific resonant states are annotated in transmission bands using the energy-angle pairs derived from the miniband analysis.

This study focuses on the first transmission band, i.e., the resonant tunneling when $E < 0.57$ eV. As the incidence angle increases from 0° to 70°, the resonant energy exhibits a progressively increasing blueshift (Fig. 2(b)). Structural parameters of the DBH can further modulate the transmission properties. Plenty of previous work have confirmed that increasing barrier width $d_{\text{InP}}$ reduces the full width at half maximum (FWHM) of the transmission resonance, while enlarging the well width $d_{\text{InAs}}$ induces a significant redshift in resonance energies[28, 34].

Under an applied bias ($V_{\text{bias}}$), the DBH potential profile becomes tilted, necessitating finer discretization in the TMM to mitigate errors from the piecewise-constant approximation. Take $V_{\text{bias}} = 0.1$ V as an example, reducing the segment width from 0.1 to 0.01 nm generates the relative error in resonance energy (first transmission band) at 0.36%, which is further lowered to 0.037% as slice width reduced to 0.001 nm. Therefore, the segment width of 0.01 nm is selected in subsequent calculation. Fig. 3(a) shows transmission spectra for normally incident electrons under bias voltages ranging from -0.1 to 0.1 V. The resonance energy shifts linearly with the bias[34]: A negative bias tilts the potential upward, inducing a blueshift, while a positive bias tilts it downward, causing a redshift. The peak value of transmission probability at zero bias reaches unity due to the coherent resonant tunneling, which will be reduced with increasing $|V_{\text{bias}}|$ because of the coherence degradation. For non-normally injected electrons, similar results can also be obtained. More importantly, the properties of the resonant tunneling are remarkably modulated by the incidence angle. For the case at $V_\text{bias} = 0.1$ V, as shown in Fig. 3(b), both resonance energy ($E_\text{peak}$) and FWHM increase with the incidence angle. The former increases from 0.1 to 0.45 eV, and the latter broadens from 0.2 to 8 meV, as $\theta$ rises from 0° to 70°.

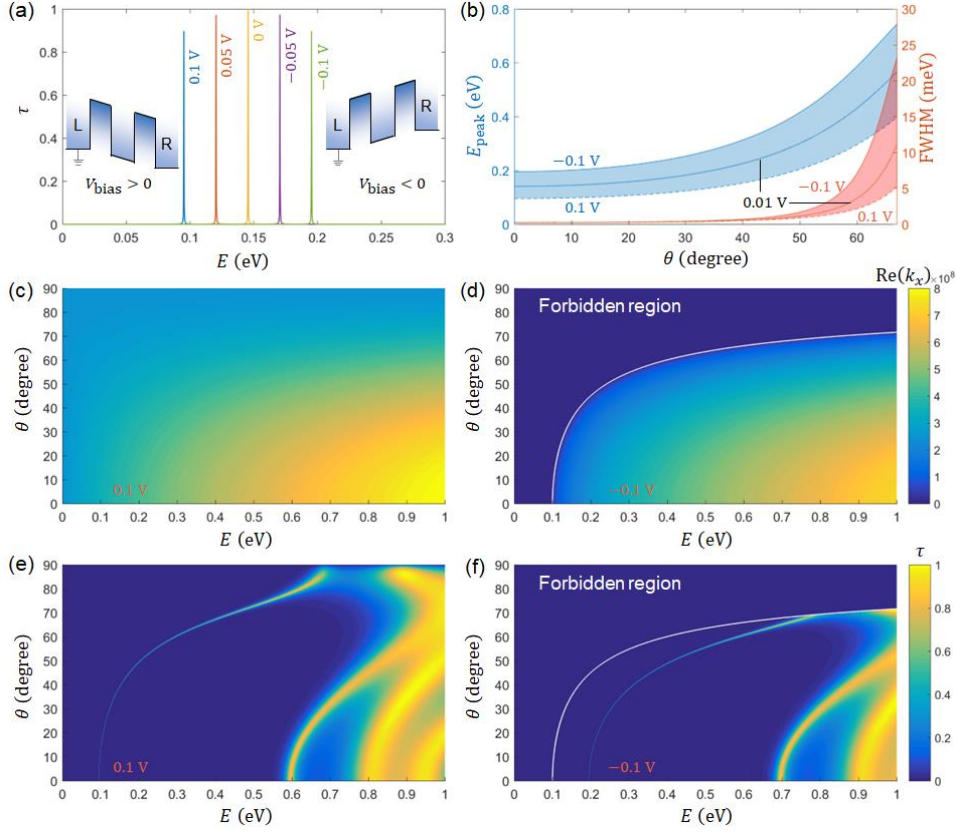

**FIG. 3** Angle-dependent electron transmission and wavevector dynamics in the DBH. (a) Transmission spectra for normally incident electrons in the energy range from 0 to 0.3 eV at different bias voltages. Insets: potential profile of DBH when $V_\text{bias} > 0$ and $V_\text{bias} < 0$, where the left electrode is grounded. (b) Angle-dependent peak energy (blue) and FWHM of transmission resonance (red) at the first transmission band for different bias voltages. (c-d) Real component of the $x$-component wavevector in the right well with respect to energy and the incidence angle for $V_\text{bias} = 0.1$ and $-0.1$ V, respectively. White line denotes $\text{Re}(k_x) = 0$, demarcating propagating and evanescent regimes. (e-f) Corresponding angularly resolved transmission spectra through the DBH.

In the biased DBH (Fig. 3(a)), the left electrode is grounded. For electrons transmitted from left to right, the wavevectors at the left and right regions are $k_\text{L} = \sqrt{2m^*_\text{InAs}E}/\hbar$ and $k_\text{R} = \sqrt{2m^*_\text{InAs}(E - qV_\text{bias})}/\hbar$, respectively, where $q$ is the electron charge. The conserved $y$-component wavevector is $k_y = k_L \sin\theta$, yielding the $x$-component in the right region: $k_x = \sqrt{k_\text{R}^2 - k_y^2}$. Successful transmission through the DBH requires the real part of $k_x$ to satisfy:

$$\text{Re}(k_x) > 0 \tag{12}$$

For $V_{\text{bias}} > 0$ (downward potential tilt), Eq. (12) always holds when $E > 0$, enabling the transmission across DBH. However, for $V_{\text{bias}} < 0$ (upward tilt), some electrons will be blocked unless Eq. (12) is workable due to the potential rise. Figs 3(c) and 3(e) show the calculated $\text{Re}(k_x)$ and the electron transmission spectrum $\tau(E,\theta)$ at $V_{\text{bias}} = -0.1$ V. Nonzero transmission ($\tau > 0$) persists for $E > 0$ except the trivial case when $\theta = 90°$. For $V_{\text{bias}} = 0.1$ V (Figs 3(d) and 3(f)), a forbidden region emerges bounded by $\text{Re}(k_x) = 0$. No electron located in the forbidden region can be transmitted through the DBH.

### III. ELECTRICAL TRANSPORT WITH THE DBH

Generally, the electrical properties of nanoscale electronic devices are typically evaluated through transfer and output characteristics, which describe the current as a function of the channel's Fermi level and the applied bias voltage, respectively. In this model, the fermi level is called chemical potential $\mu_{\text{L/R}}$, governed by the semiconductor doping level and electrode work functions, is often modulated via gate voltages. Under bias condition, $\mu_R = \mu_L + qV_{\text{bias}}$. The left and right 1D electrodes are called electron reservoirs, in which electrons obey Fermi-Dirac distribution:

$$f_{\text{L/R}} = \frac{1}{1+\exp\left(\frac{E-\mu_{\text{L/R}}}{k_B T_{\text{L/R}}}\right)} \quad (13)$$

where $k_B$ is the Boltzmann constant and $T_{\text{L/R}}$ is the temperature of the left/right reservoir. Since the dimension of DBH in *x*-component is in nanoscale, ballistic transport is dominating for electron transmission, necessitating the use of the Landauer formula to calculate the current:

$$I = \frac{2q}{h} \int_0^{+\infty} \tau(E,\theta)(f_L - f_R) dE \quad (14)$$

where the factor of "2" accounts for spin degeneracy, and $h$ the Planck's constant. The current is clearly determined by the overlap of the transmission resonance and the *Fermi window* ($\Delta f = f_L - f_R \neq 0$, defines the energy range of contributing electrons) in energy dimension. At fixed DBH geometry, $\tau(E,\theta)$ depends on $V_{\text{bias}}$ and $\theta$, while $\Delta f$ is shaped by $\mu_{\text{L/R}}$ and $T_{\text{L/R}}$ of two reservoirs.

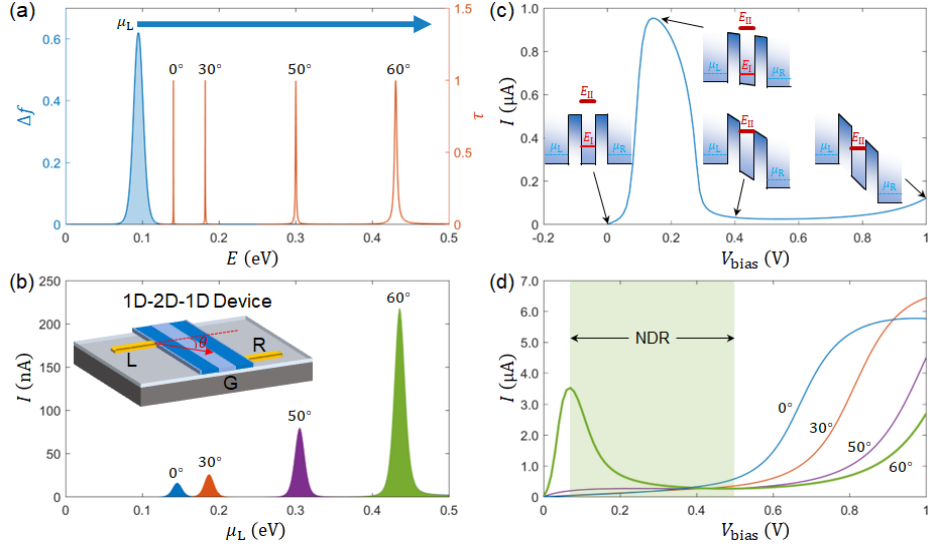

**FIG. 4** Angle-dependent electrical transport characteristics and NDR effect. (a) $\Delta f$ as the function of electron energy when $\mu_L = 0.1$ eV and $V_{bias} = 0.01$ V (blue line), and transmission spectra near the first transmission band for the incidence angles $\theta = 0°, 30°, 50°$, and $60°$ (red lines). (b) Current versus $\mu_L$ under the parameters in (a). Inset: schematic of the DBH device. (c) NDR effect observed in the output curve when $\mu_L = 0.1$ eV and $\theta = 0°$, and here $d_{InP} = 2$ nm is chosen for the presentation. Insets: band diagrams at different bias voltages. (d) Angle-dependent NDR effect as $\mu_L = 0.4$ eV.

Fig. 4(a) shows the energy-dependent *Fermi window* under the fixed parameters: $\mu_L = 0.1$ eV, $T_L = T_R = 40$ K, and $V_{bias} = 0.01$ V. This *Fermi window* with fixed shape shifts along the energy axis at varying $\mu_L$ which is mainly modulated via gate voltages in experiments (see device schematic in Fig. 4(b), inset). The transmission spectrum of the DBH exhibits pronounced angular dependence, and Fig. 4(a) shows the transmission resonances at $\theta = 0°, 30°, 50°$, and $60°$. As $\mu_L$ increases, the *Fermi window* will align with these transmission resonances, enabling measurable current and defining the transfer curve. The angular dependence of the transfer curve reveals that current magnitudes peak at maximum *Fermi window*-transmission resonance overlaps, as shown in Fig. 4(b). Notably, maximum current increases with $\theta$, a trend attributed to the broadening of transmission resonance linewidths (i.e., FWHM in Fig. 3(b)).

It is well known that the output characteristics of the DBH exhibit the typical NDR effect, arising from bias-dependent shifts in transmission resonances, as displayed in Figs. 3(a) and 3(b). Fig. 4(c) illustrates the NDR mechanism for a DBH with $d_{InP} = 2$ nm, chosen to enhance clarity. At $V_{bias} = 0$ V, the resonance energy $E_I$ of the first transmission band lies slightly above $\mu_L$, resulting in no current. As $V_{bias}$ increases, both $\mu_R$ and transmission band shift downward, driving a gradual current rise that peaks at the maximum overlap of *Fermi window* and first transmission resonance. Further bias increases push $E_\text{I}$ into the DBH bandgap, suppressing electron transmission (or current) and inducing the NDR regime. Subsequent alignment of the peak energy $E_\text{II}$ of the second transmission band with $\mu_\text{L}$ restores the current gradually. Crucially, NDR effect requires $E_\text{I}$ to exceed $\mu_\text{L}$ at zero bias and traverse the bandgap at large bias. Take $\mu_\text{L} = 0.4$ eV as an instance, $E_\text{I} = 0.146$ eV is much smaller than $\mu_\text{L}$ and no NDR effect occurs within the range of $0 < V_\text{bias} < 1$ V, as shown the case at $\theta = 0°$ in Fig. 4(d). However, angle-dependent transmission alters this behavior: increasing the incidence angle $\theta$ elevates $E_\text{I}$ (Fig. 3(b)), enabling NDR emergence. At $\theta = 60°$, for example, $E_\text{I} = 0.453$ eV is slightly larger than $\mu_\text{L}$, and generates a pronounced NDR region, as shown in Fig. 4(d). This angle-dependent NDR effect provides a viable experimental route to probe the predicted angle-dependent transmission dynamics.

## IV. ANGLE-DEPENDENT THERMOELECTRIC ENERGY MANAGEMENT

Thermoelectric effect describes the interconversion of thermal and electrical energy driven by temperature ($\Delta T$) and chemical potential ($\Delta \mu$) gradients. In our model (Fig. 5(a)), two electrodes are immersed in spaces with different temperatures ($T_\text{L} < T_\text{R}$, named as cold and hot reservoirs, respectively), and the case $\mu_\text{L} > \mu_\text{R}$ under positive bias ($V_\text{bias} = 0.01$ V) is considered. Electron transport between reservoirs is mediated by energy-selective transmission through the DBH, enabling thermoelectric power generation or heat pumping depending on the energy-dependent electron flux. In the *Fermi window*, as shown in Fig. 5(b), electrons with $E < E_0$ tends to flow from cold to hot reservoir, while from hot to cold reservoir as $E > E_0$, where $E_0 = \frac{\mu_\text{L} T_\text{R} - \mu_\text{R} T_\text{L}}{T_\text{R} - T_\text{L}}$ (derived from $f_\text{L} = f_\text{R}$). Without considering the scattering with phonons[35] and the contact resistance[36], heat flux arises from the energy carried by electron flux. For the rigorous transmission with $\delta(E)$-function, each transmitted electron in the cooling regime ($\mu_\text{L} < E < E_0$) extracts an amount of energy $E - \mu_\text{L}$ from the cold reservoir and releases $E - \mu_\text{R}$ into the hot reservoir, requiring power input $\mu_\text{c} - \mu_\text{h}$. Conversely, each transmitted electron in the power generation regime ($E > E_0$) harvests an amount of energy $E - \mu_\text{R}$ from the hot reservoir and deposits the $E - \mu_\text{L}$ into the cold reservoir, generating power output $\mu_\text{c} - \mu_\text{h}$. Both efficiencies approach Carnot limits as $E \to E_0$. Sub-$\mu_\text{L}$ electron transmission ($E < \mu_\text{L}$) invariably heats the cold reservoir due to the depletion of cold electrons. However, the hot reservoir is heated if $\mu_\text{R} <$ $E < \mu_L$ and cooled as $E < \mu_R$. These regimes are demarcated in the *Fermi window* of Fig. 5(b) (where $\mu_L = 0.11$ eV, $V_{bias} = 0.01$ V, $T_L = 40$ K, and $T_R = 80$ K), highlighting the critical role of energy-selective transmission in optimizing thermoelectric performance.

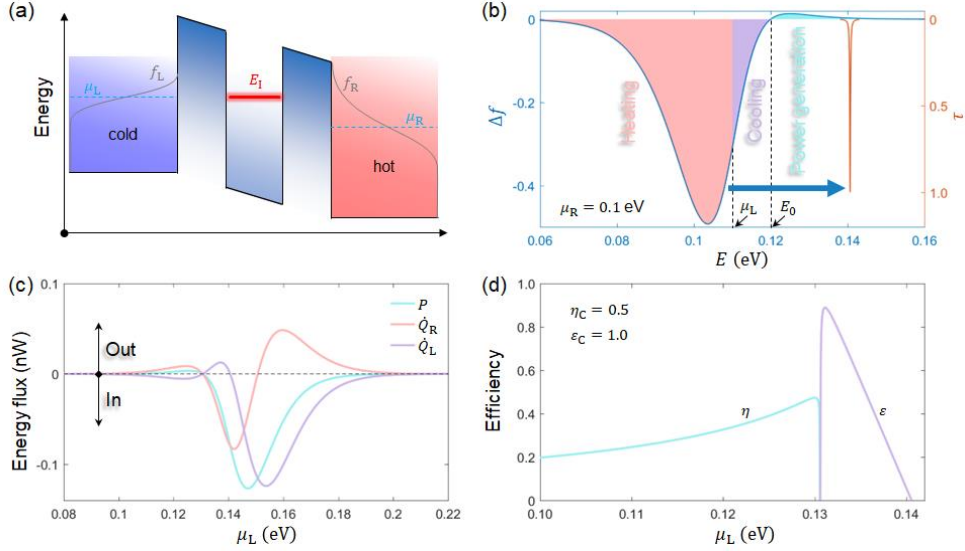

**FIG. 5** Thermoelectric energy management with the DBH for normal incidence. (a) Energy diagram of the thermoelectric device, where $T_L < T_R$ and $\mu_L > \mu_R$. Resonance energy $E_1$ at first transmission band governs energy-selective electron transport. (b) Fermi window (blue line) and the transmission spectrum around $E_1$ (red line) when $\mu_L = 0.11$ eV, $V_{bias} = 0.01$ V, $T_L = 40$ K, and $T_R = 80$ K. Electron flux in three energy regimes ($E < \mu_L$, $\mu_L < E < E_0$, and $E > E_0$) enable for cold reservoir heating, cold reservoir cooling, and power generation, respectively. The fixed-shape Fermi window shifts as $\mu_L$ varies. (c) Energy fluxes with respect to $\mu_L$, where positive values indicate power output or heat release, and negative values imply power input or heat absorption. (d) Thermoelectric efficiencies for power generation ($\eta$) and cooling ($\varepsilon$), where $\eta_C$ and $\varepsilon_C$ denote the Carnot limits at the given temperatures.

Considering the angle-dependent transmission of DBH, the heat flux per unit time and area extracted from of the cold/hot reservoir is expressed as:

$$\dot{Q}_{L/R} = \pm \frac{2}{h} \int_0^{+\infty} \tau(E,\theta)(E - \mu_{L/R})(f_L - f_R) dE \quad (15)$$

The total entropy production rate, governed by the second law of thermodynamics, is: $\Delta s = -\frac{\dot{Q}_L}{T_L} - \frac{\dot{Q}_R}{T_R}$. For transmitted electrons in an infinitesimal energy range $\delta E$, the entropy production rate becomes: $\delta s \sim \frac{(E-E_0)(T_L-T_R)}{T_L T_R}(f_L - f_R)\delta E$, yielding $\delta s \geq 0$. This ensures global adherence to $\Delta s \geq 0$ with equality only at $E = E_0$, which is consistent with the second law.

For heat engine operation (power generation mode), the power output and the efficiency are calculated by

$$P = \dot{Q}_R - \dot{Q}_L \text{ and } \eta = \frac{P}{\dot{Q}_R} \quad (16)$$

respectively, and the former can also be expressed as $P = IV_{\text{bias}}$. For refrigerator operation (cooling mode), the cooling power and the cooling efficiency are determined by

$$R = \dot{Q}_L \text{ and } \varepsilon = \frac{\dot{Q}_L}{-P} \quad (17)$$

respectively, with $-P$ denoting the power input required to sustain cooling.

Normal incidence in electron transport is often considered in conventional thermoelectric devices. For a fixed DBH geometry, thermoelectric energy management is commonly investigated by modulating the chemical potential, realized via gate voltage in experiments. Under constant bias, the *Fermi window* retains its shape but shifts along the energy axis with $\mu_L$, as depicted by the blue arrow in Fig. 5(b). Variations in the overlap between *Fermi window* and transmission resonance drive changes in energy fluxes ($\dot{Q}_L$, $\dot{Q}_R$, and $P$). Fig. 5(c) shows the calculated energy fluxes as $\mu_L$ scans from 0.08 to 0.22 eV. Distinct operational regimes emerge: including power generation ($P > 0$) for $\mu_L < 0.1306$ eV, cold reservoir cooling ($\dot{Q}_L > 0$) for $0.1306 < \mu_L < 0.1406$ eV, and cold reservoir heating ($\dot{Q}_L < 0$) for $\mu_L > 0.1406$ eV. Corresponding efficiencies for cooling and power generation are displayed in Fig. 5(d), with the maximum values of $\eta = 0.8889$ (88.9% of Carnot limit $\eta_C$) and $\varepsilon = 0.4736$ (94.7% of Carnot limit $\varepsilon_C$), enabled by the narrow transmission resonance linewidth (BWHM ~ 0.2 meV is much smaller than $k_B T$)[37, 38]. However, broad chemical potential tuning is experimentally constrained in some semiconductors because of degenerate doping[39], Fermi-level pinning[40], screening effect[41], and so on, implying that the capacity of thermoelectric energy management in Fig. 5(c) cannot be realized via gate voltage. This limitation can be addressed by shifting transmission resonances energetically through angularly resolved electron incidence.

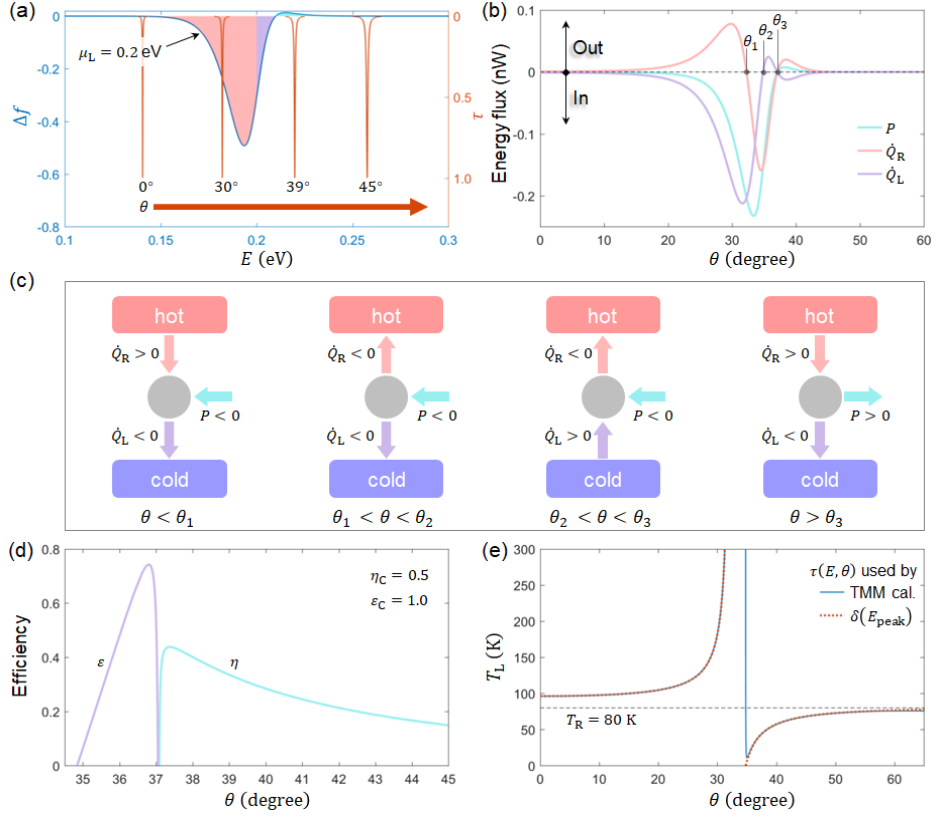

**FIG. 6** Angle-dependent thermoelectric energy management with the DBH. (a) *Fermi window* (left) and transmission spectra near the first transmission band (right) when $\mu_L = 0.2$ eV, $V_{bias} = 0.01$ V, $T_L = 40$ K, and $T_R = 80$ K. $E_{peak}$ and FWHM of transmission resonance increase with $\theta$. (b) Energy fluxes with respect to $\theta$, where $\theta_1 = 32.3°$, $\theta_2 = 34.8°$, and $\theta_3 = 37.1°$. (c) Four different regimes of thermoelectric energy management demarcated by $\theta_1$, $\theta_2$, and $\theta_3$. (d) Corresponding efficiencies for power generation ($\eta$) and cooling ($\varepsilon$). (e) Angle-dependent temperature modulation of the cold reservoir ($T_L$) at fixed $T_R = 80$ K.

As shown in Fig. 3(b), the peak energy of transmission resonance shifts by ~ 0.4 eV as the incidence angle increases from $\theta = 0°$ to 70°, enabling dynamic tuning of the overlap between the transmission resonance and the fixed *Fermi window*. Fig. 6(a) shows the *Fermi window* when the chemical potential is fixed at $\mu_L = 0.2$ eV, and the transmission resonances at $\theta = 0°, 30°, 39°$, and 45°, respectively. As $\theta$ increases, the transmission resonance passes successively through cold reservoir heating, cold reservoir cooling, and power generation regimes. After calculating, we confirmed the flexible modulation of thermoelectric energy management with respect to incidence angle, as shown in Fig. 6(b). Three critical angles are identified to demarcate different thermoelectric energy management regions, including $\theta_1 = 32.3°$, $\theta_2 = 34.8°$, and $\theta_3 = 37.1°$. The behaviors of energy fluxes at different $\theta$ are summarized in Fig. 6(c). In region I ($\theta < \theta_1$), the power input converts into heat, and then flows into the cold reservoir along with the heat flux out of the hot reservoir, i.e., $|\dot{Q}_L| = |\dot{Q}_R| + |P|$. That means the cold reservoir is heated and the hot reservoir is cooled. In region II ($\theta_1 < \theta < \theta_2$), the power input is converted into heat, and then flows into two reservoirs (i.e., $|P| = |\dot{Q}_L| + |\dot{Q}_R|$), implying that both hot and cold reservoirs are heated. In region III ($\theta_2 < \theta < \theta_3$), the cold reservoir is cooled with the help of the power input, while the hot reservoir is heated (i.e., $|\dot{Q}_R| = |\dot{Q}_L| + |P|$). In region IV ($\theta > \theta_3$), heat flows from hot to cold reservoir, with an amount of heat being converted into power output (i.e., $|\dot{Q}_R| = |\dot{Q}_L| + |P|$, the cold reservoir at present is heated, while the hot reservoir is cooled). The corresponding efficiencies for cooling (region III) and power generation (region IV) are calculated and displayed in Fig. 6(d), with maximum values at 0.7424 and 0.4383, respectively. These two values fall below those in Fig. 5(d) due to resonance broadening (FWHM increase with $\theta$), which reduces energy selectivity and enhances dissipative losses.

When the cold reservoir is thermally isolated and the right electrode is still immersed in the hot reservoir with maintained temperature $T_R = 80$ K, the thermal energy of cold reservoir can be manipulated via angular modulation of electron incidence. At thermal equilibrium state, $T_L$ can be manually prescribed as required. For the rigorous transmission with $\delta(E - E_{peak})$-function, the equilibrium condition is $\Delta f(E_{peak}) = 0$ and $T_L = \frac{E_{peak} - \mu_L}{E_{peak} - \mu_R} T_R$. It is found that $T_L$ increases from about 80 K to infinity when $E_{peak}$ increases to $\mu_R$. As $E_{peak}$ continually increases from $\mu_L$ to a large value, $T_L$ increases from 0 to about 80 K. Noticed that these temperature bounds cannot be reached both in theoretical and in experimental systems, due to transmission resonance broadening, heat conduction, thermal radiation, and so on. With the $E_{peak}$ data of Fig. 3(b) at $V_{bias} = 0.01$ V, $T_L$ as a function of $\theta$ has been obtained, as shown in Fig. 6(e), with critical incidence angles 32.2° ($E_{peak} = \mu_R$) and 34.8° ($E_{peak} = \mu_L$). For the transmission of DBH, $\dot{Q}_L = 0$ is considered as the equilibrium condition of the thermally isolated cold reservoir. The numerically calculated $T_L$ agree well with those for $\delta(E - E_{peak})$-function transmission, especially when $\theta < 32.2°$, because of the finite FWHM of the transmission resonance. For larger $\theta$, however, observable deviation is induced by the FWHM broadening. The lowest temperature for DBH is 10.9 K, which can be further lowered by reducing the FWHM of the transmission resonance.

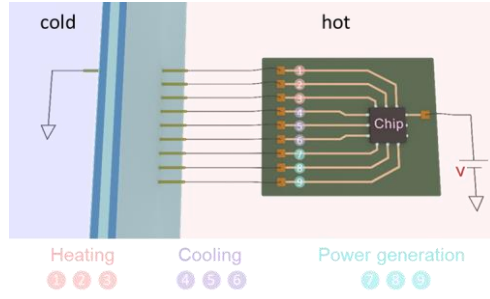

**FIG. 7** The concept of on-chip cryogenic cooling for a confined cold reservoir, different legs in 1D-2D-1D perform different thermoelectric energy management depending on the incidence angle.

## V. CONCLUSION

In this study, we proposed a 1D-2D-1D system, comprising a 2D InP/InAs/InP DBH and 1D ultrafine metallic nanowires, to enable angle-dependent electron transmission and thermoelectric energy management. By employing the TMM and low-dimensional Landauer formulism, we calculated the electronic band structure, transmission spectra, current, and heat fluxes. The results reveal resonant tunneling dynamics with a ~0.4 eV shift in resonance energy at the given electrical parameters as the incidence angle increases from 0° to 70°. Such angular tunability facilitates selective overlap with the *Fermi window*, driving transitions among power generation, cooling, and hybrid heating modes, alongside a pronounced angle-dependent NDR effect. Significantly, by circumventing material limitations (e.g., Fermi-level pinning, doping constraints), angular modulation offers a practical route to high-efficiency thermoelectric devices. For instance, we demonstrated a conceptual on-chip cryogenic cooling system (Fig. 7), where angular control of electron transmission dynamically regulates the temperature of a confined cold reservoir. These findings highlight the promise of momentum-engineered electron transmission for advancing nanoelectronics and energy-efficient thermal management systems.

**ACKNOWLEDGMENTS**

This work was supported by the Natural Science Basic Research Program of Shaanxi (Program No. 2025JC-YBMS-654) and the Natural Science Research Project of Anhui Educational Committee (Nos. KJ2021A0865 and KJ2021ZD0109)